\documentclass[prl,twocolumn,aps]{revtex4}
\usepackage{graphicx}

\begin{document}

\author{I.M. Sokolov}
\affiliation{%
Institut f\"{u}r Physik, Humboldt-Universit\"{a}t zu Berlin,
Invalidenstr. 110, D-10115 Berlin, Germany}
\title{Cyclization of a Polymer: A First Passage Problem for a Non-Markovian Process}
\date{\today}

\begin{abstract}
We discuss a problem of cyclization of a polymer molecule, which is an
important example of reaction in a system showing strongly non-Markovian
behavior on the timescales of interest. We show that the knowledge of the
joint three-time probability distribution of the end-to-end distance is
sufficient for the full description of the cyclization kinetics, so that the
survival probability follows rigorously as a solution of the Volterra
integral equation. The corresponding kinetics can easily be evaluated
numerically. We moreover discuss how do some well-known approximations
appear from this exact scheme due to decoupling.
\end{abstract}


\maketitle

Kinetics of reactions involving polymers (especially the luminescent
energy transfer) has attracted much theoretical attention in the last decade
(see e.g. \cite{Lum1,Lum2}), due to experimental relevance and to a
plenitude of applications. Cyclization of a polymer, or luminescence
quenching in a system of a donor and acceptor attached to the ends of the
polymer molecule belongs to the most complex problems in the field: It is a
case of a diffusion-controlled reaction in a system exhibiting considerably
non-Markovian behavior on the timescales of interest \cite
{Lum2,WF1,Doi,SSS,DeGennes,PZS,BG1}. This non-Markovian behavior mirrors the
essentially many-particle nature of the system. Even for the simplest
situation, in which the polymer is modelled by a Rouse chain, i.e. the
excluded volume effects and the hydrodynamical interactions between the
monomers (''beads'') are neglected, the existing theoretical approaches fail
to give an exact description of the situation and make additional physical
assumptions. Thus, their domain of applicability is inevitably restricted.
In what follows we give an exact solution for a special situation when the
reaction takes place on a contact, and show that it only depends on
three-time joint probability density of the end-to-end distance. We also
discuss how do some standard approximations emerge.

The theoretical approaches typically start from the Langevin equation for
the whole chain of $N\gg 1$ monomers, and try to reduce the overall dynamics
to the manageable effective one. The simplest approach consists in
describing the evolution of the end-to-end distance $r$ as an effective
diffusion \cite{SSS}. A more accurate Wilemski-Fixman theory (WF) \cite{WF1} 
and its equivalents \cite{Doi,DeGennes} express the reaction rates through the
effective sink-sink correlation function and assume local equilibration. The
comparison of theoretical results with the results of numerical simulations
of the underlying dynamics \cite{PZS} for the case when the reaction takes
place with probability 1 whenever the ends of the chain (the monomers 0 and $%
N$) approach each other at the distance $a$, shows that the WF approach
leads to systematic overestimate of the mean first contact time of the ends
of the polymer, while neglecting the effective diffusion assumption leads to
underestimates. The simulations in \cite{PZS} started from the equilibrium
(Boltzmann) initial distribution of the end-to-end distance $r_{0}$, with $%
r_{0}>a$. Another approach \cite{BG1} is based on the reduction of the full
equation of motion to the equation of motion of the end-to-end distance
using the projection operator formalism. The corresponding equation is again
approximate. We note that the methods of Refs. \cite
{WF1,Doi,DeGennes,BG1} are based on the
Smoluchowski approach in the theory of diffusion-controlled reactions, in
which the reaction probability follows from the solution of the full or
reduced diffusion equation.

In the case of the reaction taking place with probability 1 on contact, as
discussed in Ref.\cite{PZS}, introducing a sink term in the exact or reduced
Fokker-Planck equation (like in the WF theory or in Ref. \cite{BG1}) is
equivalent to posing an absorbing boundary condition $P(r-a)=0$ (a ''black sphere''). Such problem can be reformulated as the first passage problem
from the initial distance $r=r_{0}$ between the ends to $r=a$ 
\cite{SSS,PZS}. In the
Markovian case, such first passage problem can be solved using an
alternative approach \cite{Redner} based on the renewal property; the idea
stemming from Montroll and Weiss \cite{MW}. This approach can be generalized
to the non-Markovian situation.

In what follows we concentrate on the solution of the initial-condition
problem, not averaged over the initial positions. The reason is two-fold:
First, the recent success of micromanipulation of polymers makes the
situation experimentally relevant, while no theoretical predictions for the
kinetic curves are available. Second,
the initial condition problem renders clear the \textit{geometrical} nature
of standard approximations \cite{WF1,Doi,DeGennes}. The results for the 
equilibrium initial condition can be obtained by additional numerical 
averaging.

Let us first note that the reaction process depends only on the single
variable $r$ so that the problem
is essentially one-dimensional. Moreover, the trajectories of our process
are continuous and nowhere differentiable, just like the trajectories of the
usual Wiener process are (\textit{vide infra}). Let us consider the relation
between the distribution of the first passage time to a sphere of the radius 
$r=a$ around the origin of the coordinates $F(a,t\mid r_{0},0)$ and the
conditional probability for the ends to be found at distance $r$: If the
trajectory starts at point $r_{0}$, and is found at $r$ at time $t$, it may
have already visited $r$ before, at some time $t^{\prime }<t$. Thus, the
conditional probability to be at $r$ at time $t$, provided the particle
started at $r_{0}$ at $t=0$, $G(r,t\mid r_{0},t_{0})$, is given by
the following equation: 
\begin{eqnarray}
G(r,t &\mid &r_{0},0)=\delta (r-r_{0})\delta (t)  \label{Gen} \\
+ &&\int_{0}^{t}F(r,t^{\prime }\mid r_{0},0)G(r,t\mid r,t^{\prime
};r_{0},0)dt^{\prime },  \nonumber
\end{eqnarray}
where $G(r,t\mid r,t^{\prime };r_{0},0)$ is the conditional probability to
be at $r$ at time $t$, provided $r$ was visited earlier at time $t^{\prime }$
and that the particle started at $r_{0}$ at $t=0$. This equation is
essentially the definition of the first passage time distribution and holds
for all processes with continuous trajectories, whether Markovian or not.
For a Markovian process the conditional probability $G(r,t\mid r_{0},0)$ is
a Green's function, and $G(r,t\mid r,t^{\prime };r_{0},0)$ depends only on
the latest the arguments to the right of the line, so that $G(r,t\mid
r,t^{\prime };r_{0},0)\equiv G(r,t\mid r,t^{\prime })$. In this case, our
Eq.(\ref{Gen}) reduces to a well-known renewal equation for the
first-passage time \cite{Redner}. If the particle definitely does not start at $r_{0}=r$, the $%
\delta $-functional term can be omitted.

According to the Bayes formula, $G(r,t\mid
r_{0},0)=P(r,t;r_{0},0)/P(r_{0},0)$ and $G(r,t\mid r,t^{\prime
};r_{0},0)=P(r,t;r,t^{\prime };r_{0},0)/P(r^{\prime },t^{\prime };r_{0},0)$
where $P(r_{0},0)$, $P(r,t;r_{0},0)$ and $P(r,t;r,t^{\prime };r_{0},0)$ are
the one-, two- and tree-time joint probability distributions. The
corresponding one-dimensional distributions as functions of $r$ can be
expressed as integrals of the overall joint probabilities over the
surface of the sphere of radius $r$, so that $P(r,t\mid
r_{0},t_{0})=\int_{S}d\mathbf{s}P(\mathbf{r},t;\mathbf{r}_{0},t_{0})$ and $%
P(r,t;r,t^{\prime };r_{0},t_{0})=\int_{S}\int_{S}d\mathbf{s}d\mathbf{s}%
^{\prime }P(\mathbf{r},t;\mathbf{r},t^{\prime };\mathbf{r}_{0},t_{0})$, so
that the overall equation reads 
\begin{eqnarray}
\frac{\int_{S}d\mathbf{s}P(\mathbf{r},t;\mathbf{r}_{0},0)}{P(\mathbf{r}%
_{0},0)} &=&  \label{Rigorous} \\
\int_{0}^{t}F(r,t^{\prime } &\mid &r_{0},0)\frac{\int_{S}\int_{S}d\mathbf{s}d%
\mathbf{s}^{\prime }P(\mathbf{r},t;\mathbf{r},t^{\prime };\mathbf{r}_{0},0)}{%
\int_{S}d\mathbf{s}^{\prime }P(\mathbf{r}^{\prime },t^{\prime };\mathbf{r}%
_{0},0)}dt^{\prime },  \nonumber
\end{eqnarray}
(where $\left| \mathbf{r}_{0}\right| >r$). It is reasonable to rewrite this
equation in the following form: 
\begin{equation}
\int_{0}^{t}F(r,t^{\prime }\mid r_{0},0)Q(t,t^{\prime },r_{0})dt^{\prime }=1,
\label{Volterra}
\end{equation}
with the kernel 
\begin{equation}
Q(t,t^{\prime },r_{0})=\frac{\int_{S}\int_{S}d\mathbf{s}d\mathbf{s}^{\prime
}P(\mathbf{r},t;\mathbf{r},t^{\prime };\mathbf{r}_{0},0)P(\mathbf{r}_{0},0)}{%
\left[ \int_{S}d\mathbf{s}^{\prime }P(\mathbf{r}^{\prime },t^{\prime };%
\mathbf{r}_{0},0)\right] \left[ \int_{S}d\mathbf{s}P(\mathbf{r},t;\mathbf{r}%
_{0},0)\right] }.  \label{Kernel}
\end{equation}
This is an exact equation expressing the pdf of first passage time through
the three-time joint probability distribution of the end-to-end distance. As
stated, we concentrate on the initial value problem and discuss the
distribution of the first passage times as depending on $r_{0}$.

Note that for the Rouse chain and many other so-called generalized Gaussian
structures \cite{SB} the random process describing temporal
changes of the distance between each two beads is Gaussian, since it is a
weighted sum of many Gaussian random variables describing the uncorrelated 
displacements of beads. Thus, for example, the
three-time joint probability distribution is 
\begin{equation}
P(\mathbf{r},t;\mathbf{r},t^{\prime };\mathbf{r}_{0},0)=\frac{(2\pi )^{-9/2}%
}{\sqrt{\det \mathbf{\hat{A}}}}\exp \left( -\frac{1}{2}\mathbf{R\hat{A}}^{-1}%
\mathbf{R}\right) ,
\end{equation}
where the vector $\mathbf{R}$ is a 9-component vector $(x,y,z,x^{\prime
},y^{\prime },z^{\prime },x_{0},y_{0},z_{0})$ and the covariance matrix $%
\mathbf{\hat{A}=}\left[ \left\langle R_{i}R_{j}\right\rangle \right] $
consists of 9 diagonal blocks: 
\begin{equation}
\mathbf{\hat{A}=}\left( 
\begin{array}{lll}
\mathbf{D}(0) & \mathbf{D}(t-t^{\prime }) & \mathbf{D}(t) \\ 
\mathbf{D}(t-t^{\prime }) & \mathbf{D}(0) & \mathbf{D}(t^{\prime }) \\ 
\mathbf{D}(t) & \mathbf{D}(t^{\prime }) & \mathbf{D}(0)
\end{array}
\right)
\end{equation}
where $\mathbf{\hat{D}}(t)$ is a diagonal matrix $\mathbf{\hat{D}}(t)=%
\mathbf{\hat{I}}\phi (t)$ ($\mathbf{\hat{I}}$ is a unit matrix). The
function $\phi (t)$ is a relaxation function of the structure, $\phi
(t)=\left\langle x(t)x(0)\right\rangle $. The corresponding two-point
distributions read 
\begin{equation}
P(\mathbf{r},t;\mathbf{r}_{0},0)=\frac{(2\pi )^{-3}}{\sqrt{\det \mathbf{\hat{%
B}}}}\exp \left( -\frac{1}{2}\mathbf{R\hat{B}}^{-1}\mathbf{R}\right)
\end{equation}
with 
\begin{equation}
\mathbf{\hat{B}=}\left( 
\begin{array}{ll}
\mathbf{D}(0) & \mathbf{D}(t) \\ 
\mathbf{D}(t) & \mathbf{D}(0)
\end{array}
\right) ,
\end{equation}
and the one-point density reads $P(\mathbf{r}_{0},0)=\left[ 2\pi \phi
(0)\right] ^{-3/2}\exp \left[ -\left( r_{0}^{2}/2\phi (0)\right) \right] .$

The integrals over the two-point functions, say $\int_{S}d\mathbf{s}P(%
\mathbf{r},t;\mathbf{r}_{0},0)$, can be evaluated analytically. Taking the $%
z $-axis to follow the direction of $\mathbf{r}_{0}$ one gets (r=a): 
\begin{eqnarray}
&&\int_{S}d\mathbf{s}P(\mathbf{r},t;\mathbf{r}_{0},0)=  \label{P2} \\
&&\frac{2a^{2}}{\pi }\frac{\exp \left[ -\frac{\phi (0)(a^{2}+r_{0}^{2})}{%
2\left[ \phi ^{2}(0)-\phi ^{2}(t)\right] }\right] \sinh \left[ \frac{%
ar_{0}\phi (t)}{\left[ \phi ^{2}(0)-\phi ^{2}(t)\right] }\right] }{%
ar_{0}\phi (t)\sqrt{\phi ^{2}(0)-\phi ^{2}(t)}}.  \nonumber
\end{eqnarray}

Let us now consider the three-time pdf $\int_{S}\int_{S}d\mathbf{s}d\mathbf{s%
}^{\prime }P(\mathbf{r},t;\mathbf{r},t^{\prime };\mathbf{r}_{0},0)$. Taking
the point $\mathbf{r}_{0}$ to lay on the $z$-axis and the end-point 
$\mathbf{r}$ to have a
zero $y$-coordinate, the integrals over the azimuthal angles can be
performed analytically, so that the corresponding distribution reduces to 
\begin{eqnarray}
&&P(r,t;r,t^{\prime };r_{0},t_{0})=\frac{a^{4}}{(2\pi )^{5/2}A^{3/2}\{\phi \}%
}\times   \label{P3} \\
&&\exp \left[ -\frac{a^{2}(2\phi _{0}^{2}-\phi _{2}-\phi
_{3})+r_{0}^{2}(\phi _{0}^{2}-\phi _{1}^{2})}{2A\{\phi \}}\right] \times  
\nonumber \\
&&\times \int_{0}^{\pi }\int_{0}^{\pi }I_{0}\left[ \frac{a^{2}\sin \psi \sin
\vartheta (\phi _{0}\phi _{1}-\phi _{2}\phi _{3})}{A\{\phi \}}\right]  
\nonumber \\
&&\times \exp \left[ f(a,r_{0},\{\phi \})/A({\psi})\right] \sin \psi \sin \vartheta
d\psi d\vartheta ,  \nonumber
\end{eqnarray}
where $\vartheta $ and $\psi $ are the corresponding polar angles. Here  
$A\{\phi \}=\phi _{0}^{3}+2\phi _{1}\phi _{2}\phi _{3}-\phi _{0}(\phi
_{1}^{2}+\phi _{2}^{2}+\phi _{3}^{2})$, $I_0$ is the modified Bessel function,
and
\begin{eqnarray}
&&f(a,r_{0},\{\phi \})=a^{2}\cos \psi \cos
\vartheta (\phi _{2}\phi _{3}-\phi _{0}\phi _{1})+  \nonumber \\
&&+ ar_{0}\left[ (\phi _{1}\phi _{2}-\phi _{0}\phi _{3})\cos \vartheta
+(\phi _{1}\phi _{3}-\phi _{0}\phi _{2})\cos \psi \right]  
\end{eqnarray}
where $\phi _{0}=\phi (0)$, $\phi _{1}=\phi (t-t^{\prime })$, $\phi
_{2}=\phi (t)$ and $\phi _{3}=\phi (t^{\prime })$. These forms are universal
and apply to any Gaussian model, whether linear or branched chain, or a
network.

Let us now turn to a special case of the Rouse chain. For long enough chain
the time correlation function $\phi (t)$ can be approximated through 
\begin{equation}
\phi (t)=\left\langle L^{2}\right\rangle \frac{8}{\pi ^{2}}\sum p^{-2}\exp
(-\lambda _{p}t)
\end{equation}
where for a Rouse chain the summation runs over the odd integers $p$ and $%
\lambda _{p}=p^{2}/\tau _{R}$ \cite{DoiEdw}. In what follows we use the rms 
size of the chain 
$\left\langle L^{2}\right\rangle ^{1/2}$ and the Rouse time $\tau _{R}$ as
length and time units, so that $r$ and $t$ are nondimensional. In these units
we get $\phi (t)=\frac{8}{\pi ^{2}}%
\sum_{n=0}^{\infty }(2n+1)\exp (-(2n+1)^{2}t)$. This function is continuous
in zero, so that $\left\langle \left[ x(t)-x(0)\right] ^{2}\right\rangle
=2\left[ 1-\phi (t)\right] $ tends to zero for small $t$, which fact
verifies the continuity of the trajectories necessary for our consideration.
The Volterra equation can be then solved numerically by approximating the
integral in Eq.(\ref{Volterra}) by a finite sum. This solution and the
evaluation of the double integral, Eq.(\ref{P3}) in the kernel is performed
using MATHCAD. The corresponding first-passage time distributions for
different initial conditions are shown in Fig.1 for the initial distances $%
r_{0}=1,2,4$ and 8 and for $a=0.1$. The numerical accuracy of the results is
around one per cent. We note that with decreasing the distance, the typical
mean first passage time decreases very fast; the maximum of the curve
corresponding to $r_{0}=1$ is at $t_{\max }\sim 0.01$, and is not
resolved on the scales of Fig.1. This effect is due to the small-scale
motion corresponding to higher modes \cite{Doi,DeGennes} and mirrors the
compact exploration of space by the chain's end. On the other hand, the
large-scale relaxation is mostly connected with the lowest mode, which is
slow.

\begin{figure}
\includegraphics{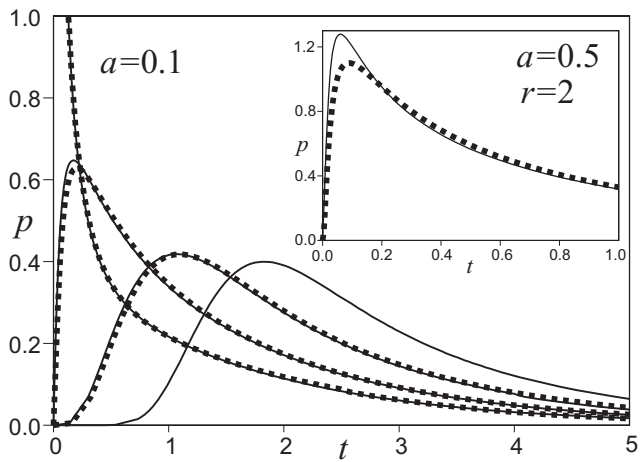}
\caption{The first passage time pdf's for $a=0.1$ and $r_{0}=1,2,4$ and 8
(from left to right). The solid curves correspond to the solution of
equation, Eq.(\ref{Volterra}) with the exact kernel, Eq.(\ref{Kernel}). The
dotted lines (shown for $r_{0}=1,2,4$) correspond to the Wilemski-Fixman
approximation, Eq.(\ref{QWF}), see text for details. The inset shows the
case $a=0.5$ and $r_{0}=2$, for which the WF-approximation ceases to perform
reasonably.}
\end{figure}

Let us now discuss the emergence of the WF-approximation and the Markovian
approximation of Ref. \cite{SSS}  within
our scheme and return to Eq.(\ref{Rigorous}). The WF approximation
corresponds to a pseudo-Markovian decoupling $G(r,t\mid r,t^{\prime
};r_{0},0)\simeq G(r,t\mid r,t^{\prime })$ in Eq.(\ref{Gen}), which, in the
language of $P$ is: $P(\mathbf{r},t;\mathbf{r},t^{\prime };\mathbf{r}%
_{0},0)\simeq P(\mathbf{r},t;\mathbf{r},t^{\prime })P(\mathbf{r}_{0},0)$ and $P(%
\mathbf{r}^{\prime },t^{\prime };\mathbf{r}_{0},0)\simeq P(\mathbf{r}^{\prime
},t^{\prime })P(\mathbf{r}_{0},0)$. This corresponds to the approximation of
the kernel $Q$ by 
\begin{equation}
Q_{WF}(t,t^{\prime },r_{0})=\frac{\int_{S}\int_{S}d\mathbf{s}d\mathbf{s}%
^{\prime }P(\mathbf{r},t;\mathbf{r},t^{\prime })P(\mathbf{r}_{0},0)}{\left[
\int_{S}d\mathbf{s}^{\prime }P(\mathbf{r}^{\prime },t^{\prime })\right]
\left[ \int_{S}d\mathbf{s}P(\mathbf{r},t;\mathbf{r}_{0},0)\right] }.
\label{QWF}
\end{equation}
Note that this approximation has a purely geometrical nature: It assumes
that the distance from the point $\mathbf{r}_{0}$ to all points on the
surface $S$ is approximately the same and corresponds to using the mean
value theorem in evaluating the integrals in the r.h.s. of Eq.(\ref{Rigorous}%
). This assumption is exact for $r_{0}=0$ and for $r_{0}\rightarrow \infty $%
; in our case $r_{0}>a$ it delivers a low-order expression in $a/r_{0}$. The
solutions of Eq.(\ref{Volterra}) with the kernel $Q_{WF}$ are shown in Fig.1
for $r_{0}=1,2$ and $4$ with dashed lines. We note that for $a/r_{0}$
small the approximation performs excellently and leads to slight
overestimate of the first passage times. The discrepancies grow when $a/r_{0}
$ gets larger: fixing for example, $r=2$ and increasing the value of to $%
a=0.5$ we find that the height of the maximum is underestimated by 15\% on
the cost of the somewhat fatter tail of the distribution; the mean value of
the first passage time is overestimated by around 3\%. For even larger $%
a/r_{0}$ the accuracy of the approximation worsens rapidly.

We now show how do the usual equations of the Wilemski-Fixmann theory appear
from this decoupling approximation. To do this we return to Eq.(\ref
{Rigorous}), now reading as 
\begin{eqnarray}
\frac{\int_{S}d\mathbf{s}P(\mathbf{r},t;\mathbf{r}_{0},0)}{P(\mathbf{r}%
_{0},0)} &=&  \label{WiF} \\
\int_{0}^{t}F(r,t^{\prime } &\mid &r_{0},0)\frac{\int_{S}\int_{S}d\mathbf{s}d%
\mathbf{s}^{\prime }P(\mathbf{r},t;\mathbf{r},t^{\prime })}{\int_{S}d\mathbf{%
s}^{\prime }P(\mathbf{r}^{\prime },t^{\prime })}dt^{\prime },  \nonumber
\end{eqnarray}
and average both sides over the equilibrium distribution of $r_{0}$, $%
P(r_{0},0)$. We note that the function multiplying $F$ under the integral in
the r.h.s. does not depend on $r_{0}$, so that only $F$ is averaged, giving
rise to the averaged first-passage distribution $\bar{F}$, and that the
average in the l.h.s. is a one-time marginal distribution, i.e. $\int_{S}d%
\mathbf{s}P(\mathbf{r},t)$. Thus, the equation for  $\bar{F}$ reads: 
\begin{equation}
\int_{0}^{t}\bar{F}(r,t^{\prime }\mid r_{0},0)\frac{\int_{S}\int_{S}d\mathbf{%
s}d\mathbf{s}^{\prime }P(\mathbf{r},t;\mathbf{r},t^{\prime })}{\int_{S}d%
\mathbf{s}P(\mathbf{r},t)\int_{S}d\mathbf{s}^{\prime }P(\mathbf{r}^{\prime
},t^{\prime })}dt^{\prime }=1.  \label{WiF2}
\end{equation}
Noting that the process $\mathbf{r}(t)$ is stationary we denote $%
\int_{S}\int_{S}d\mathbf{s}d\mathbf{s}^{\prime }P(\mathbf{r},t;\mathbf{r}%
,t^{\prime })=C(t-t^{\prime })$. Due to correlation decoupling at long
time, $\int_{S}d\mathbf{s}P(\mathbf{r},t)\int_{S}d\mathbf{s}^{\prime }P(%
\mathbf{r}^{\prime },t^{\prime })$ is exactly $C(\infty )=C_{\infty }$,
if the corresponding limit does not vanish. Thus, we can write 
\begin{equation}
\int_{0}^{t}\bar{F}(r,t^{\prime }\mid x_{0},0)\frac{C(t-t^{\prime })}{%
C(\infty )}dt^{\prime }=1.
\end{equation}
Applying the Laplace transform to the both sides of the equation, we get
$1/u=\tilde{F}(u)\tilde{C}(u)/C_{\infty }$, so that 
\begin{equation}
\tilde{F}(u)=C_{\infty }/\left[ uC(u)\right] .
\end{equation}
Moreover, since $%
C_{\infty }=\lim_{t\rightarrow \infty }C(t)$, for $u$ small one has $\tilde{C%
}(u)=C_{\infty }/u+A+...$, with $A=\lim_{u\rightarrow 0}(C(u)-C_{\infty }/u)$. 
Noting that the mean first passage time 
$\tau =\int_{0}^{\infty }tF(t)dt = -\frac{d}{du}\left. \tilde{F}(u)\right| _{u=0}$,
one arrives at 
\begin{eqnarray}
\tau  &=&C_{\infty }\left. \left( \frac{1}{u^{2}}\frac{1}{\tilde{C}(u)}+%
\frac{1}{u}\frac{\tilde{C}^{\prime }(u)}{\tilde{C}^{2}(u)}\right) \right|
_{u\rightarrow 0}=\frac{A}{C_{\infty }}=  \nonumber \\
&=&\int_{0}^{\infty }\left( \frac{C(t)}{C_{\infty }}-1\right) dt,
\end{eqnarray}
which is exactly Eq.(12) of Ref. \cite{PZS}. The approach of Ref. \cite{SSS}
is an approximation of the same nature: It assumes that the random process $%
r(t)$ is an Ornstein-Uhlenbeck process (the only one Gaussian Markovian
process, the one with $\phi (t)\simeq \exp (-\alpha t)$, with $\alpha
=2/\tau _{R}$  \cite{PZS} ), for which Eq.(\ref{WiF}) is exact. Compared to the WF-theory
it contains an additional assumption, namely one of the effectively
exponential relaxation of $\phi $ and has typically a lower accuracy.

Compared to the general case, our solution applies to a rather special sink
function (''black sphere'', or a $\delta$ -sink), but our derivation here is
rigorous and much less technical than the original WF approximation, and can
apply to a variety of cases where the approximation fails. Let us say a
few words on the general sink function. In general, in the WF theory one has 
$C(t-t^{\prime })=\int_{V}\int_{V}d\mathbf{r}d\mathbf{r}^{\prime }S(%
\mathbf{r})S(\mathbf{r}^{\prime })P(\mathbf{r},t;\mathbf{r},t^{\prime })$,
where $S(\mathbf{r)}$ characterizes the sink strength. Although such
expressions have no immediate counterparts within the first passage time
formalism, they can be obtained as mean-field approximations when one
associates $S(\mathbf{r)}$ with the density of absorbing regions in a
problem where the absorbing boundary is corrugated or even not
singly-connected. In this case, the corresponding surface integrals Eq.(\ref
{WiF2}) are changed for the volume integrals weighted with the sink function.
The same approximation can be done, of course, also in a general
non-Markovian form, Eqs.(\ref{Rigorous}-\ref{Kernel}). The corresponding
forms may be no more exact, but still take into account mode details of the
process than WF ones do.

Let us summarize our findings. The cyclization of a polymer 
is an example of a diffusion-controlled reaction
in a system exhibiting strongly non-Markovian behavior on the timescales of
interest. Considering the reaction as a first-passage process, one is able
to provide an exact equation governing the reaction probability. This
equation can be readily solved numerically, giving the kinetic curves. We
also discuss how the popular Wilemski-Fixman approximation emerges from the
exact scheme as a pseudo-Markovian decoupling.

The partial financial support by the Fonds der Chemical Industrie is
gratefully acknowledged.

\end{document}